\documentstyle[preprint,aps]{revtex}

\newcommand{\gra}[1]{\mbox{\boldmath $ #1 $}}
\input{epsf}
\input{psfig}

\title{Non--gaussian probability distribution functions
       in two dimensional Magnetohydrodynamic turbulence}
\author{L. Sorriso--Valvo$^1,2$, V. Carbone$^1$, P. Veltri$^1$, H.
Politano$^2$ \and A. Pouquet$^2$}
\address{$^1$ Dipartimento di Fisica, Universit\`a degli studi della Calabria,
87036 Rende (CS), Italy, \\ and Istituto Nazionale di Fisica della
Materia Unit\`a di Cosenza\\
$^2$ Laboratoire G. D. Cassini, CNRS UMR 6529, OCA Observatoire de Nice,
B. P. 4229, 06304 Nice Cedex 04, France}

\begin{document}

\maketitle

\begin{abstract}

Intermittency in MHD turbulence has been analyzed using high resolution
2D numerical simulations. We show that the
Probability Distribution Functions (PDFs) of the fluctuations of the 
Els\"asser fields, magnetic field and velocity field depend on the scale at 
hand, that is they are self--affine. The departure of the PDFs from a Gaussian
function can be described through the scaling behavior of a single parameter
$\lambda_r^2$ obtained by fitting the PDFs with a given curve stemming from 
the analysis of a multiplicative model by Castaing et al. \cite{Castaing}. 
The scaling behavior of the parameter $\lambda_r^2$ can be used to extract 
informations about the intermittency. 
A comparison of intermittency properties in different MHD turbulent flows
is also performed.

\end{abstract}
\clearpage


One of the relevant features of turbulent systems is small--scale intermittency 
\cite{Frisch}.
Starting from the Kolmogorov 1962 theory \cite{Kolmogorov}, several models 
have been developed to take into account the intermittency effects on the 
statistics of the turbulent fields. Measurements show the presence of
self--affine fields, which is the most 
remarkable of such effects, leading to the scaling departure from gaussianity 
of the Probability Distribution Functions (PDFs) of turbulent fluctuations 
(see \cite{Frisch} and references therein). The usual statistical tool 
to check such departure is the analysis  of the scaling exponents
$\zeta_p^{\psi^{\prime}}$ of the longitudinal structure  functions of an (as yet) 
unspecified vector field ${\gra {\psi^{\prime}}}$, {\it i.e.} the
moments of the distributions of the fluctuations,  namely: 
$S_r^{(p)}(\delta\psi^{\prime}_r) = \langle \delta\psi^{\prime p}_r \rangle = 
\langle [({\gra 
\psi^{\prime}}({\gra x}+{\gra r})-{\gra \psi^{\prime}}({\gra x}))\cdot{\gra r}/r]^p \rangle$; 
here ${\gra \psi^{\prime}}$ represents either the velocity field 
${\gra v}$, the normalized magnetic induction
${\gra b}={\gra B}/\sqrt{4\pi\rho}$ ($\rho$ is the constant density), 
or the Els\"asser variables
${\gra z}^{\pm} = {\gra v} \pm {\gra b}$; $r=|{\gra r}|$ is the scale we 
are examining and homogeneity is assumed. The structure 
functions are assumed to scale like 
$S_r^{(p)}(\delta \psi^{\prime}) \sim r^{\zeta_p^{\psi^{\prime}}}$
in the inertial range.
In the non--intermittent case the scaling exponents $\zeta_p^{\psi^{\prime}}$ should 
display a linear dependence on the order index of the moments; in the framework 
of the Kolmogorov theory, for example, this dependence is a 
$\zeta_p^{v}=p/3$ law. 
However, experimental observations have shown a nonlinear behavior of such
scaling exponents, suggesting the multifractal nature of
the energy transfer \cite{Frisch,Paladin}. 
This kind of behavior has been observed
both in neutral fluids and in plasmas \cite{Frisch,Biskamp}.
A different approach to the study of intermittency consists in analysing the 
scaling behavior of the PDFs of fluctuations instead of the scaling of the 
moments \cite{Frisch,Biskamp,Monin,Vincent}. 
In the framework of the multiplicative cascade, the presence of very 
intense fluctuations of the fields becomes more probable as the scale 
decreases, because of the progressive strong localization of the active 
structures. As a result, the tails of the PDFs are higher than gaussian ones, 
and rare events become significant in the statistics. 
By fitting experimental data with a model distribution function, it is possible 
to describe intermittency through a small set of parameters.
The characterization performed through the structure functions, 
on the contrary, requires, in principle, the determination of an infinite set 
of scaling exponents. 
Unfortunately, predictions and models concerning the scaling behavior of the PDFs 
are less common and less firmly established
than those concerning the scaling behavior of structure functions, a notable 
exception being the turbulence coming from Burgers equation \cite{Kraichnan}.
In this paper, we follow one such model for PDFs \cite{Castaing}, and 
we show that the informations thus obtained  
are in agreement with the previous results derived directly
from moments.

\begin{figure}[h!]
\centerline{{\hskip 3.5cm} 
\psfig{file=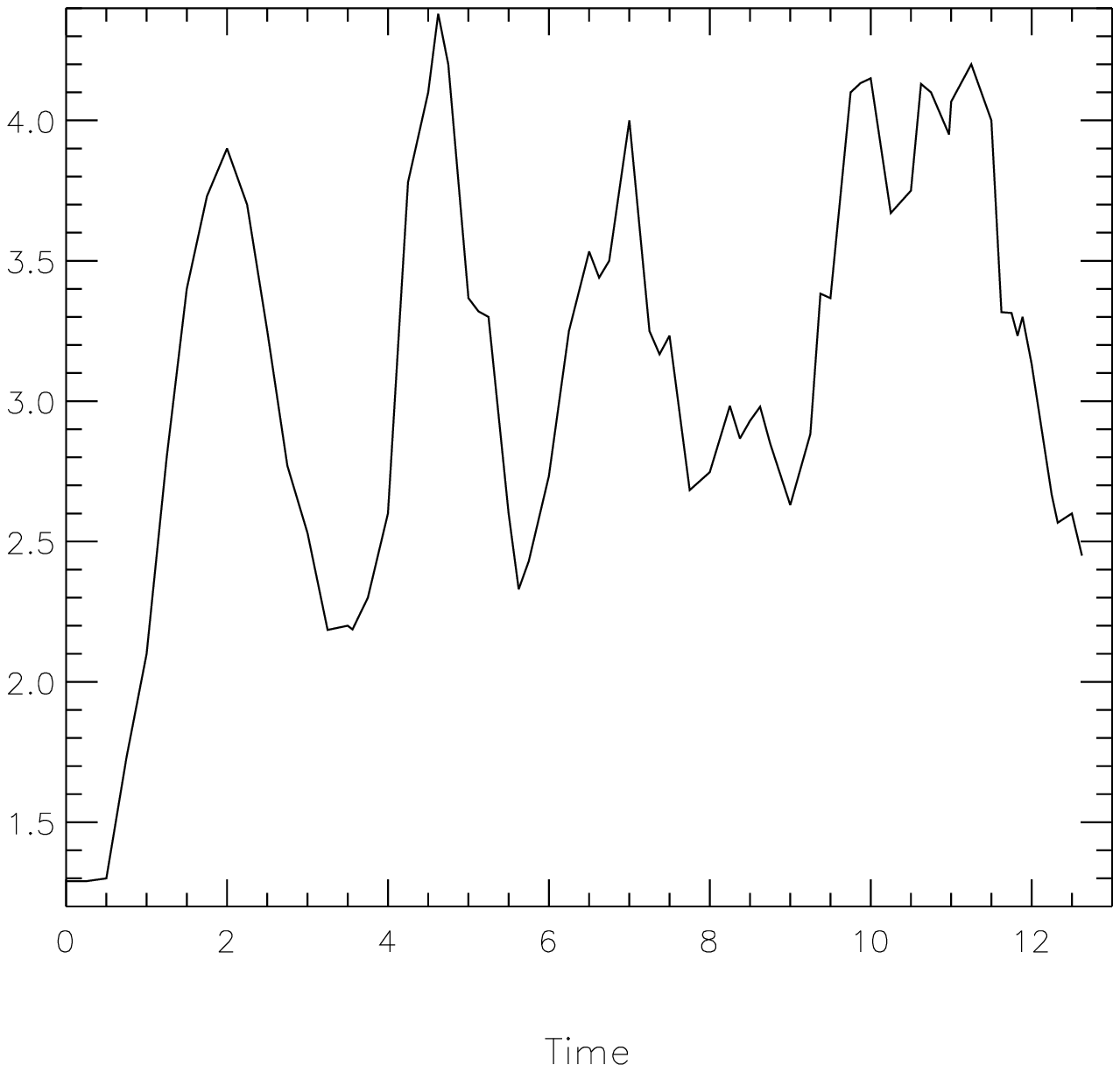,width=8.0cm}\psfig{file=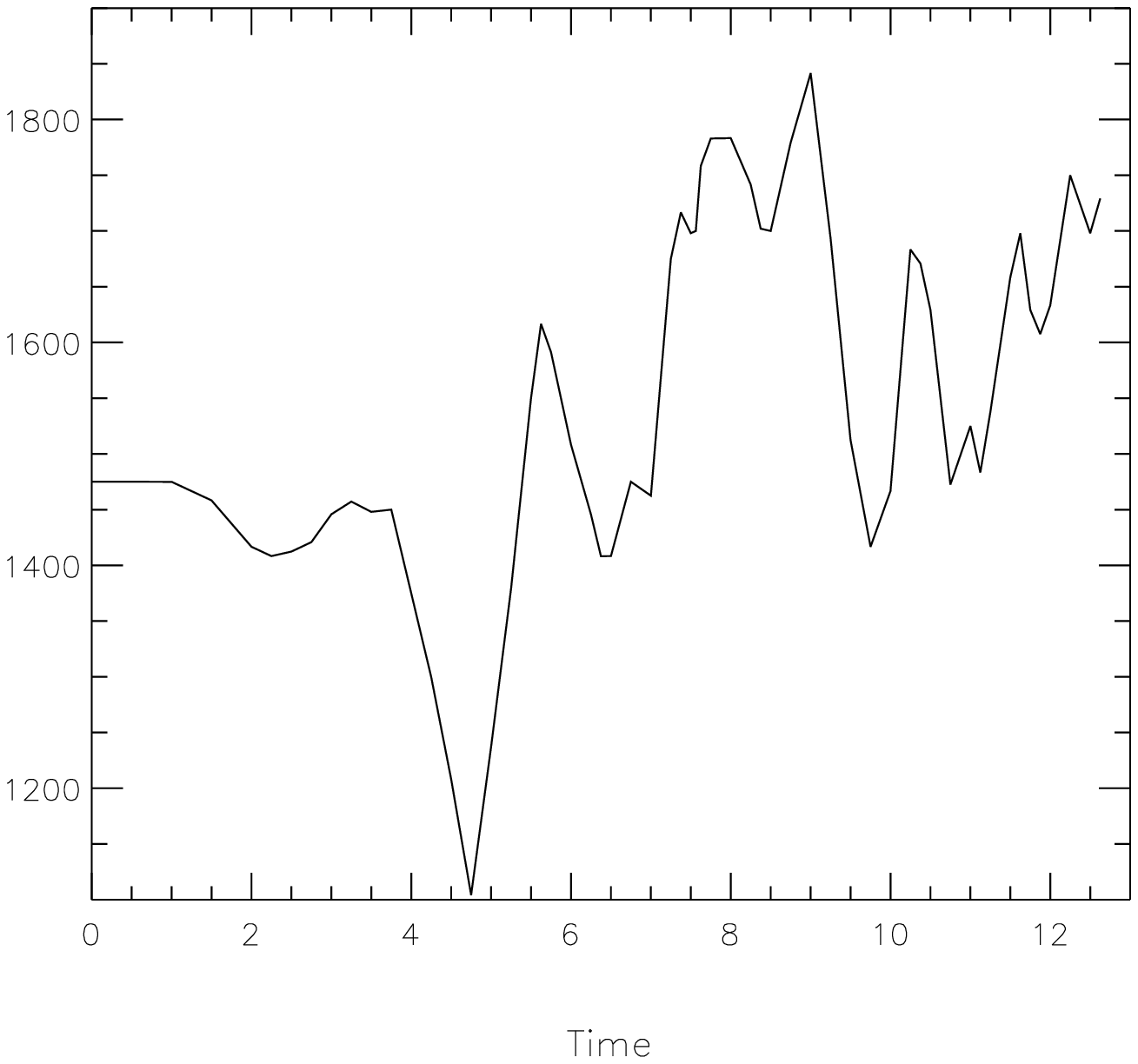,width=8.0cm}}
\vskip 1cm
\caption{On the left, temporal evolution of the ratio of magnetic to 
kinetic energy $E^M/E^V(t)$ and, on the right, 
temporal evolution of the integral Reynolds number.}
\label{tempo}
\end{figure}

The data we use stem from the turbulent fields obtained from two--dimensional 
magnetohydrodynamics incompressible simulations, with periodic boundary 
conditions at a resolution of $1024^2$ grid points 
(see also \cite{Politano}).
The forcing consists in maintaining
constant the amplitudes of Fourier modes with $|{\bf k}|=1$, and the resulting
energy is concentrated at large scales. There is on average an excess
of magnetic energy $E^M$ over its kinetic counterpart $E^V$, as can be seen
on the left of figure \ref{tempo} showing the temporal evolution of the ratio $E^M/E^V$;
better equipartition is obtained in the small scales as diagnosed by the
ratio of enstrophies (not shown).
The data analyzed in the paper are averaged over approximatively
160 eddy turnover times in the statistically steady state, 
from $t=6.30$ to $t=12$. 
This lengthy computation is rendered necessary by the long--time 
fluctuations in the flow, as is visible on figure 
\ref{tempo};
these fluctuations are induced by the large--scale forcing.
The data sets thus consist of about
$10^7$ points, and up to now, these are the largest data sets used to
investigate intermittency in statistically steady MHD flows.
The temporal window of analysis is chosen on the basis of the stationarity
of the integral Reynolds number, shown on the right of figure \ref{tempo}.

\begin{figure}[htb]
\epsfxsize=10cm    
\centerline{\epsffile{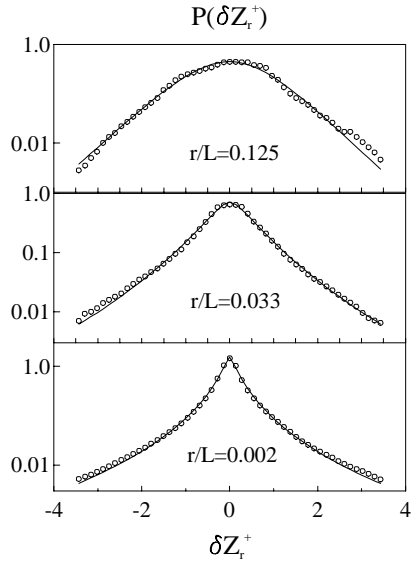}}
\caption{PDFs of the fluctuations of the normalized Els\"asser
variable ${\gra z}^+$ for three different scales (see insert). 
The full line represents the fit 
made with the convolution function (\ref{pdfs}).}
\label{pdf}
\end{figure}

\begin{figure}[htb]
\epsfxsize=9cm    
\centerline{\epsffile{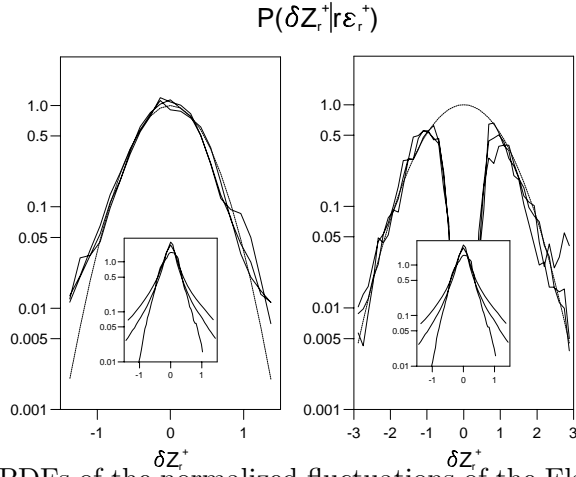}}
\vskip -12pt
\vskip -12pt
\caption{Conditioned PDFs of the normalized fluctuations of the Els\"asser 
variables ${\bf z}^+$  for three different scales,
namely $r/L=0.002$, $r/L=0.03$ and $r/L=0.25$. 
The PDFs $P(\delta z_r^+\mid r\varepsilon_r^+)$ are 
conditioned by a given level of the energy dissipation
$r\varepsilon^+_r=(\delta z_r^+)^2\delta z_r^-$ 
between $-0.1$ and $0.1$ (left panel) and between $0.9$ and $1$ 
(right panel). Gaussian fits are reported for comparison (dashed lines). 
The standard deviations of such curves, computed from the fit, are 
$\sigma_r=0.4$ and $\sigma_r=0.9$, respectively.
The insets show the unconditioned PDFs for the same values of $r/L$.}
\label{cond}
\end{figure}

In order to get PDFs with zero average and unit standard deviation, we use,
as is customary, the normalized fluctuations of the turbulent fields: 
$$\delta \psi_r = \frac{\delta \psi^{\prime}_r - 
\langle \delta \psi^{\prime}_r \rangle}{\langle (\delta \psi^{\prime}_r - 
\langle \delta \psi^{\prime}_r \rangle)^2 \rangle^{1/2}} \; ,$$
\noindent
where $\delta \psi^{\prime}_r$ is any one of the relevant fields at scale $r$. 
We built up the PDFs of the normalized fluctuations by dividing the range 
between $-3.5$ and $+3.5$ (in standard deviation units) in bins of length 
$A_{bins}$. We then used the histogram of the fluctuations 
$N_r(\delta \psi_r)$ to compute the probability density: 
$$P(\delta \psi_r) = \frac{N_r(\delta \psi_r)}{A_{bins}N_{tot}(r)} \; ,$$ 
where $N_{tot}(r)$ is the actual number of fluctuations at scale $r$ 
in the interval $(-3.5,3.5)$.
The PDFs have been computed for several values of the scale $r$, ranging from 
$r/L=10^{-3}$ up to $r/L=0.5$, where $L=2 \pi$ is the length of the 
simulation box. 
Samples of the so obtained PDFs are shown in figure \ref{pdf} for the
normalized fluctuations of the ${\bf z^+}$ field; similar PDFs hold
for the ${\bf z^-}$, magnetic and velocity fields. 
As can be easily seen, the PDFs are gaussian at large scales, 
but become more and more stretched as the scale decreases. 
This scaling behavior is present for all fields and is an indication 
that these fields are self--affine. Such an 
intermittent behavior, in the framework of  the Kolmogorov's refined
similarity hypothesis \cite{Frisch}, can be attributed to the fluctuations of
the energy transfer rate, and the  dependence on scales of the PDFs can be
eliminated by looking at the PDFs  conditioned to a given value of the energy
transfer rate at the scale $r$  (see \cite{Sreenivasan,Castaing93} for the
fluid case).  In MHD, we can introduce the local energy transfer rates for both
Els\"asser variables, defined as: 
$r\varepsilon_r^{\pm}=(\delta z_r^{\pm})^2\delta z_r^{\mp}$ (see e. g. 
\cite{Politano}). 
The PDFs $P(\delta z_r^+|r\varepsilon_r^+)$, conditioned to two different values 
of $\varepsilon_r^+$, are shown in figure \ref{cond}. 
The same gaussian shape is observed at all scales, which means that intermittency
has been eliminated. We found the same behavior for the PDFs 
$P(\delta z_r^-|r\varepsilon_r^-)$. As reported in the caption of 
figure \ref{cond}, for a fixed scale $r$, the width of the PDFs 
can be different when conditioned by different values of the energy transfer 
rate. The departure from gaussianity observed in the
unconditioned PDFs is then the imprint of the multifractal model
of intermittency. In fact for each scale $r$, the
intermittency can be described as the  superposition of different PDFs of
fluctuations belonging to subsets with  different $\varepsilon_r^{\pm}$ 
transfer rates.
Using the multifractal framework \cite{Frisch,Paladin,Benzi}, 
we can give a quantitative 
analysis of the continuous scaling departure of PDFs from a gaussian. In fact, 
as stated by Castaing et al. \cite{Castaing} in order to describe the PDFs at a 
given scale $r$, two ingredients are needed: the parent distribution at the 
scale $L$ and the distribution of the energy transfer rate. The first 
information can be extracted directly from the experimental observations, 
just looking at the large scale PDFs of the fields fluctuations which is a 
Gaussian curve. The second item needs some {\it a priori} hypothesis about the 
shape of the PDF of the $\varepsilon_r^{\pm}$ transfer rates. 
In other words, the resulting PDFs 
of fluctuations at scale $r$ can be described as a convolution of the 
parent gaussian distribution with a chosen distribution for 
$\varepsilon_r^{\pm}$.
Following \cite{Castaing}, this can be done in a continuous way by introducing
a distribution ${\cal L}_{\lambda_r}(\sigma_r)$ for the width $\sigma_r$ of the 
gaussians (which is directely proportional to $\varepsilon_r^{\pm}$), and by 
then computing the convolution:

\begin{equation}
P(\delta \psi_r) = 
\int_0^{\infty} 
{\cal L}_{\lambda_r}(\sigma_r)\exp \left (-\delta \psi_r^2/ 2 \sigma_r^2 \right)
{1 \over \sqrt{2 \pi} \sigma_r} 
{d\sigma_r \over \sigma_r} \ .
\label{pdfs}
\end{equation}

In this paper, we use the log--normal ansatz, as in Castaing et 
al. \cite{Castaing}:
 
\begin{equation}
 \ \  {\cal L}_{\lambda_r}(\sigma_r) = {1 \over \sqrt{2 \pi} \lambda_r}
\exp\left(-{\ln^2 \sigma_r/\sigma_{0,r} \over 2 \lambda_r^2}
\right) \ .
\label{lognormal}
\end{equation}
The PDF of a field is thus seen, in this model, as 
a superposition of gaussian curves 
whose standard deviations are distributed according to a log--normal law.
The parameter $\sigma_{0,r}$ is the most probable value of the  
$\sigma_r$ deviations for a given scale $r$,
while the parameter $\lambda_r$ represents the width of the 
log--normal distribution ${\cal L}_{\lambda_r}(\sigma_r)$.
In so doing, the scaling properties of the PDFs are concentrated into the single
scaling behavior of the parameter $\lambda^2_r$. For $\lambda_r^2=0$, the 
distribution ${\cal L}_{\lambda_r}(\sigma_r)$ is a delta function, and
the convolution 
(\ref{pdfs}) is a single gaussian of width $\sigma_{0,r}$. As $\lambda_r^2$ 
increases, the spectrum of the values of $\sigma_r$, involved in the 
convolution, is wider and the tails of the PDFs consequently become stronger.

\begin{figure}[htb]
\epsfxsize=9cm     
\centerline{\epsffile{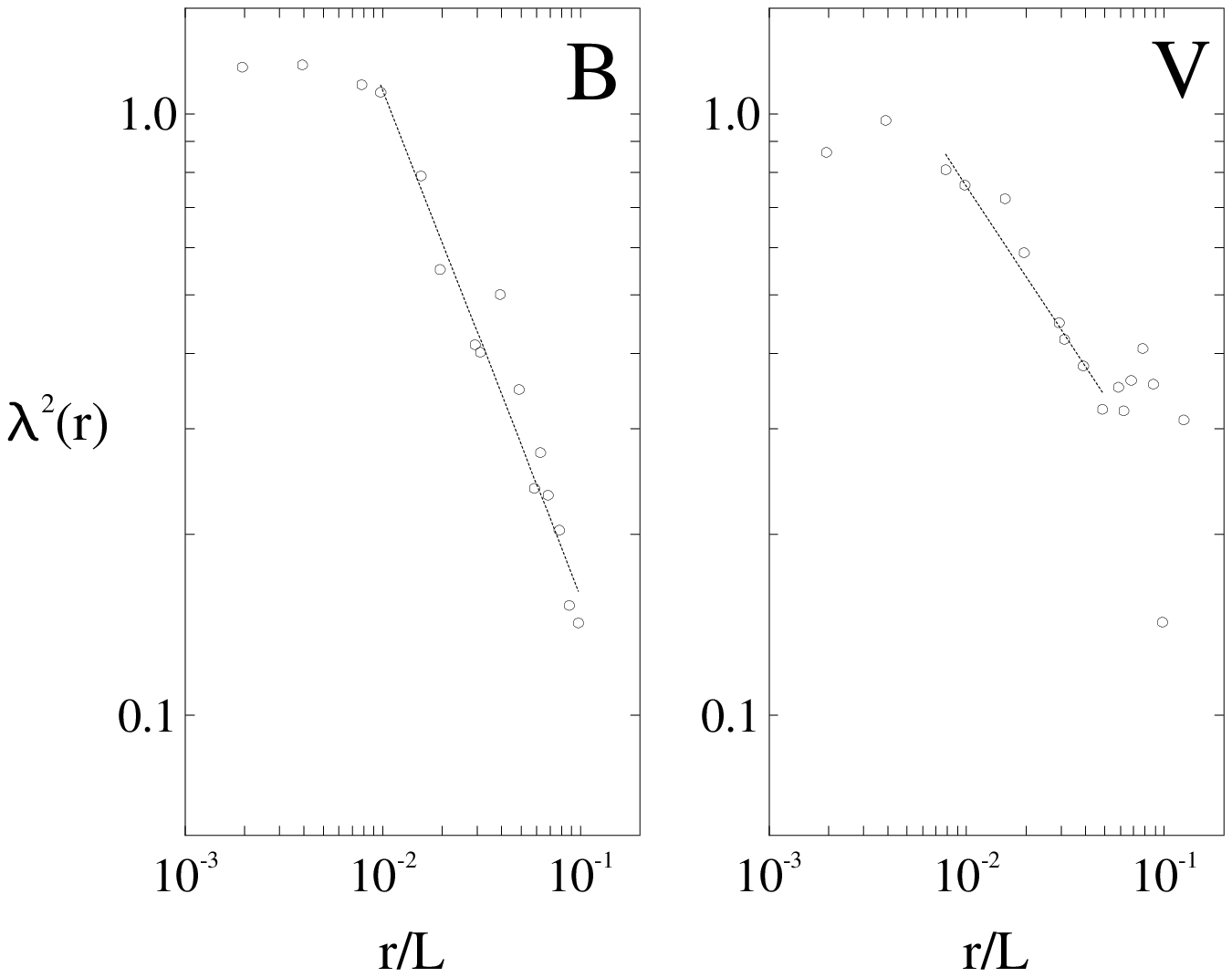}}
\vskip -12pt
\epsfxsize=9cm     
\centerline{\epsffile{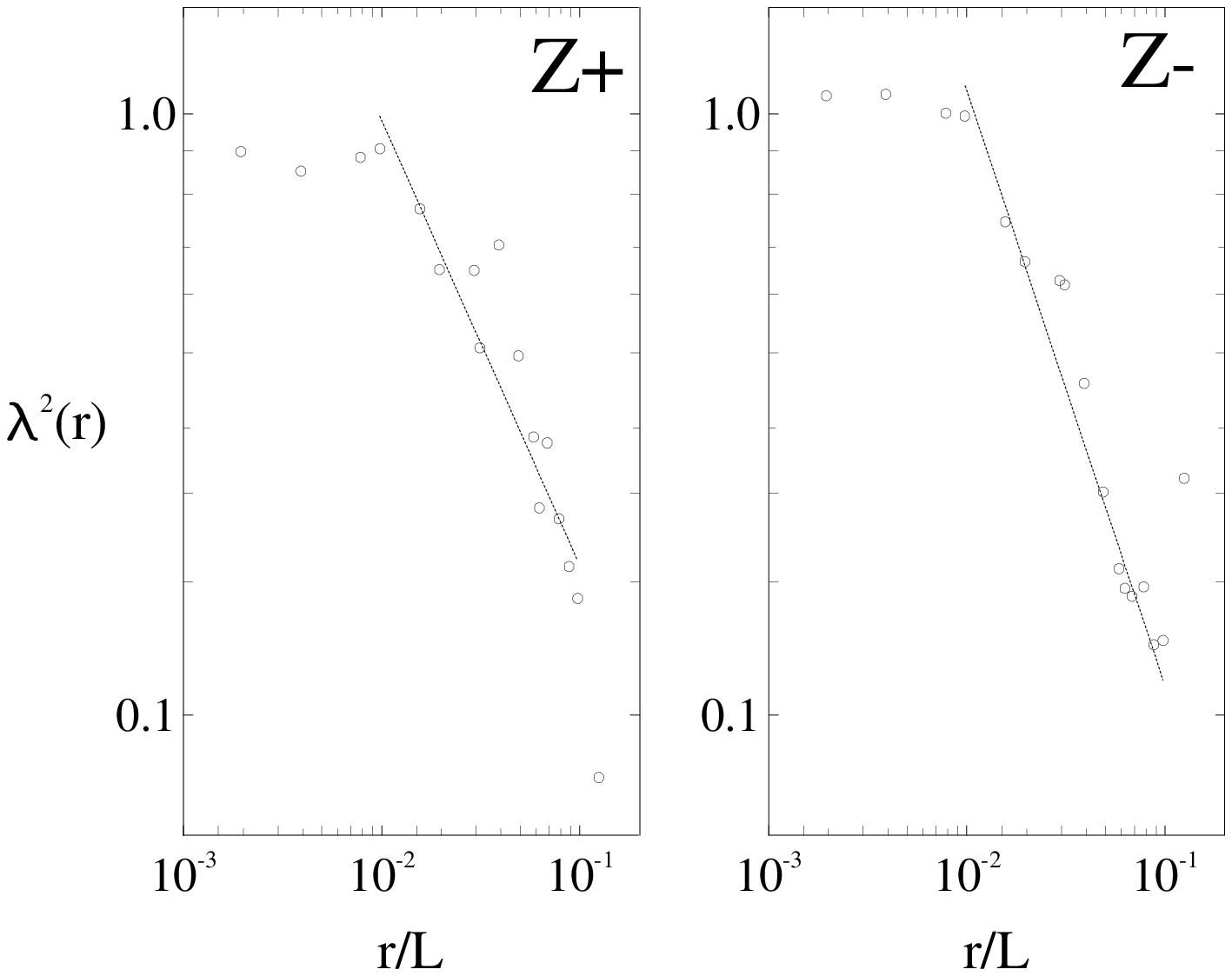}}
\caption{Scaling behavior of the exponent $\lambda^2_r$ for the 
magnetic field, the velocity and the Els\"asser variables (see insert). 
Straight lines represent the fit with power-laws as reported in Table 
\ref{table1}.}
\label{lam2}
\end{figure}

The results of the fit of the measured PDFs with the model (\ref{pdfs}) are 
shown in figure \ref{pdf}. It can be seen that the model 
reproduces quite well the scaling properties of the PDFs $P(\delta \psi_r)$
of the various fields, although with a lesser agreement at the 
largest scales of the flow. 
In order to get informations about intermittency, we compute $\lambda_r^2$
from the PDFs using (\ref{pdfs})-(\ref{lognormal})
 and we then look at the scaling of this parameter 
with the separation length $r/L$ (figure \ref{lam2}). 
When considering the scaling of the structure functions
$S^{(p)}_r(\delta {\bf z}^{\pm})$, it was shown
in \cite{Politano} that the inertial scales for the Els\"asser variables, 
fundamental in this problem \cite{grl}, 
are in the domain $0.01 \le r/L \le 0.1$.
This can be our main guide for scaling here. 
We however perform a careful examination of the data to check the scaling 
ranges of $\lambda^2_r$;
in particular, for the velocity field, this will lead us to shorten this range at large
scales (see table \ref{tableone}).
A typical nontrivial behavior can be distinguished as a 
power--law scaling:

\begin{equation}
\lambda_r^2(r) \sim (r/L)^{-\beta} \; ,
\label{lambda2}
\end{equation}
in the previously defined inertial range:
a saturation of $\lambda_r^2$ is 
reached at the onset of the dissipative range, while the gaussian 
regime is found on scales larger than $r/L \simeq 0.1$. 

This method leads to a characterisation of the turbulent system using only 
two parameters: the scaling exponent $\beta$ and $\lambda^2_{max}$, 
the maximum value of $\lambda^2_r$ over $r$.
The parameter $\beta=-d\log\lambda_r^2/d \log(r/L)$, 
the logarithmic derivative of $\lambda_r^2$,
represents how ``fast'' the generation of intermittency occurs
throughout the inertial range. In fact, a greater value of $\beta$ is related 
to an energy cascade mechanism which produces intermittency in a more 
efficient way (the wings of the distribution of $\delta \psi_r$ increase more
quickly). The parameter $\lambda^2_{max}$, being related to both the maximum 
number of gaussians needed in the convolution (\ref{pdfs}) and to the weight of 
the widest gaussians in that convolution, tells us how ``strong'' is the 
intermittency, {\it i.e.} how deeply the intermittent cascade is active 
and generates the strongest events $\delta \psi_r$. 

The computed values of $\lambda_r^2$ provide $\lambda^2_{max}$ and
$\beta$ which are given in table \ref{tableone}, together with their variations
obtained by a chi--square test, as well as the range 
of scales where (\ref{lambda2}) is verified. Note that the analysis of each
temporal data set separately led us to eliminate some samples: 
$\lambda_r^2$ was ill--defined on average on one third of the individual
temporal data sets.
The magnetic field appears more intermittent than the velocity, 
in agreement with previous numerical results \cite{Politano,Gomez},  
and with the results in solar wind plasmas \cite{Sorriso} 
(see also \cite{biskamp} for a 3D computation in a
decaying helical MHD flow using hyper viscosity). This result resembles what
happens in fluid turbulence
where passive scalars are more intermittent than the velocity field
\cite{Ciliberto}. In MHD the stronger intermittency of the magnetic field is
perhaps due to the fact that, when nonlinearly coupled with a velocity field,
the magnetic field behaves like a passive vector, at least in
the kinematic phase of the dynamo.
The Els\"asser variables (dominated by the magnetic field, see figure 
\ref{tempo}) display also a strong intermittency;
the values of the parameters found for the ${\bf z^-}$ field
are comparable to those for the magnetic field, while the parameters
found for ${\bf z^+}$ are intermediate between those obtained for the
magnetic field and the velocity.
The slight difference between the $\pm$ $\beta$--parameter could 
appear significant: 
although the global correlation coefficient between the velocity and the 
magnetic field is low when averaged over the whole flow, it presents 
regions with strong pointwise values of either signs (not shown) 
and this may reflect in the $\pm$ discrepancy observed here. 
A similar result is found in the solar wind 
\cite{privatecommbyLuca}.

As a comparison with some previous experimental analysis, the values 
of $\beta$ obtained here are larger than those found for fluid flows 
($\beta \simeq 0.3$) \cite{Castaing}, 
for solar wind turbulence ($\beta \simeq 0.2$) \cite{Sorriso} 
and for magnetic turbulence in a Reverse Field Pinch laboratory 
plasma ($\beta \simeq 0.4$) \cite{Carbone99}. 
This fact can be interpreted as an indication that 2D MHD turbulence in the  
present numerical simulations is strong, with a low degree 
of anisotropies and inhomogeneities which could exist, on the contrary, in  
geophysical or laboratory plasma turbulence. 
Again,  the values of
$\lambda^2_{max}$ found here are greater than those found in the solar wind
(slightly so for the magnetic field, more clearly for the velocity) 
\cite{privatecommbyLuca}, and the
magnetic field is strongly more intermittent than in the laboratory plasma
mentioned above.

In conclusion, this paper shows that the analysis of the PDFs is a useful tool
to quantify intermittency. In the present approach, only two parameters,
$\beta$ and $\lambda^2_{max}$,
are used to characterize the intermittency in a way which is
found consistent with previous analysis. 
An improvement of the method using different distributions for the energy transfer rates
is presently in progress.

\acknowledgments

Numerical simulations were performed at IDRIS (Orsay).
We are thankful to a referee for a useful remark.
H. Politano and A. Pouquet aknowledge partial financial 
support from CNRS, PNST.

\begin{table}
\caption[Table]{For the different fields, values of the two parameters $\beta$ 
and $\lambda_{max}^2$, together with their statistical error bars, determined 
from the indicated ranges of fit (see text).
}
{\label{table1}}
\begin{center}
\begin{tabular}{cccc}
&&&\\[-20pt]
      &     $\beta$    & $\lambda ^2_{max}$ &       range of fit      \\
\hline&&&\\[-4pt]
$v$   & $0.50 \pm 0.15$ & $0.8 \pm 0.2$ & $0.008 \le r/L \le 0.05 $ \\
$b$   & $0.84 \pm 0.13$ & $1.1 \pm 0.3$ & $0.01  \le r/L \le 0.1  $ \\
$z^+$ & $0.74 \pm 0.14$ & $0.9 \pm 0.2$ & $0.01  \le r/L \le 0.1  $ \\
$z^-$ & $0.98 \pm 0.19$ & $1.0 \pm 0.3$ & $0.01  \le r/L \le 0.1  $ \\
[2pt]\end{tabular}
\label{tableone}
\end{center}

\end{table}

\end{document}